\documentstyle[aps,manuscript,epsf]{revtex}

\input epsf

\def\to{\rightarrow}

\def\be{\begin{equation}}
\def\ee{\end{equation}}
\def\bea{\begin{eqnarray}}
\def\eea{\end{eqnarray}}

\begin{document}

\title{Muon Anomalous Magnetic Moment and \\Lepton Flavor Violation
}

\author{T. Huang, Z.-H. Lin, L.-Y. Shan and X. Zhang\\}

\address{
Institute of High Energy Physics, Chinese Academy of Sciences,
\\ P.O. Box 918, Beijing, 100039, China}

\maketitle

\begin{abstract}
A non-universal interaction, which involves only the third family leptons
induces lepton flavor violating couplings and contributes to the anomalous
magnetic moment of muon. In this paper, we study the effects of
non-universal interaction on muon (g-2) and rare decay $\tau \to \mu \gamma$
by using an effective lagrangian technique, and
a phenomenological $Z^\prime$ model where $Z^\prime$ couples only to the third
family lepton. We find that the deviation from the theory can be explained and
the induced $\tau \to \mu \gamma$ rate could be very close to
the current experimental limit. In the $Z^\prime$ model, $M_{Z^\prime}$
has to be lighter than 2.6 TeV.
%\vspace{.5cm}
\end{abstract}

Recently the Brookhaven AGS experiment 821 group has announced a new measurement
of the anomalous magnetic moment with three times higher accuracy than it was
previously known~\cite{E821}. The experimental data now is in conflict with
the theoretical prediction of the Standard Model (SM) with an excess of 2.6 $\sigma$.
This may be an indication of new physics if it is more than a statistical
fluctuation. Many authors have considered various possibilities of
interpreting the discrepancy in models beyond the Standard Model,
such as, supersymmetric version of the standard model, muon with substructure
in composite model and anomalous gauge boson couplings~\cite{new}.

In this paper we consider a class of models where non-universal interactions
exist and study the effects of new physics on muon (g-2) and lepton rare decay.
We first take a model independent approach (effective lagrangian) to new
physics, then consider a phenomenological $Z^\prime$ model. We find that
the non-universal interaction can account for the deviation of muon (g-2) from
the SM prediction. Furthermore, our models predict the rare decay rate
$\tau \to \mu \gamma$ which can be tested in the planned $\tau$-charm factories
and in $Z^\prime$ model $M_{Z^\prime}$ has to be lighter than 2.6 TeV which
results in many interesting phenomenon~\cite{nonuni,lan}.

The tau lepton is the heaviest one in the lepton sector and especially
the top quark is heavier than all other fermions, with mass close to the
electroweak symmetry breaking scale, the new physics, if exists, is expected to become
manifest in the effective interaction of the third family fermions~\cite{peccei}.
There are also theoretical and phenomenological motivations that there may
exist interactions with generation non-universal couplings~\cite{nonuni,lan}.
Furthermore, the possible anomalies in the $Z$- pole $b \bar{b}$ asymmetries
may suggest a non-universal $Z^\prime$~\cite{erler,chan}.

In terms of an effective lagrangian, the new physics is described by higher dimensional
operators.
%In Ref.~\cite{tau} we have argued that dimension-six operators involving
%only the tau lepton induce an anomalous
%magnetic moment of the muon and lepton flavor violating decay $\tau \to \mu
%\gamma$, etc. We will in this paper point out that to fit the muon (g-2) to
%the new data within 90\% C.L., $\tau \to \mu \gamma$ rate is predicted to be
%very close to the current experimental limit. However, the predicted tau (g-2)
%is still several orders of magnitude below the current experimental limit.
At dimension-six, there are two operators \cite{tau,esc} which contribute directly to the
anomalous magnetic moment of leptons,
\begin{eqnarray}\label{ope}
{\cal O}_{\tau B}&=&{\bar L}\sigma^{\mu\nu} \tau_R \Phi B_{\mu\nu}, \nonumber\\
{\cal O}_{\tau W}&=&{\bar L} \sigma^{\mu\nu} {\vec \sigma} \tau_R \Phi {\vec
W}_{\mu\nu},
\end{eqnarray}
where $L=( \nu_{\tau}, ~ \tau_L )$, $\Phi$ is the Higgs scalar,
$B_{\mu\nu}$ and $W_{\mu\nu}$ are field strengths of $U_Y(1)$ and $SU_L(2)$,
$\vec \sigma$ are Pauli matrices.

In the presence of operators ${\cal O}_{\tau B}$ and ${\cal O}_{\tau W}$,
the effective Lagrangian ${\cal L}_{eff}$ can be written as,
\begin{eqnarray}\label{eq5}
{\cal L}_{eff} = {\cal L}_0 + \frac{1}{\Lambda^2} ( C_{\tau B} {\cal O}_{\tau B}
    + C_{\tau W} {\cal O}_{\tau W} + h.c. ),
\end{eqnarray}
where ${\cal L}_0$ is the standard model lagrangian and $C_{\tau B}$,
$C_{\tau W}$ are constants which represent the coupling strengths of
${\cal O}_{\tau B}$ and ${\cal O}_{\tau W}$. $C_{\tau B}$, $C_{\tau W}$
are expected to be $O(\frac{1}{(4\pi)^2})$ in theories with weakly interaction
~\cite{wudka}, however could be large if the fundamental theory is strongly
interaction such as the models of composite tau lepton.

After the electroweak symmetry is broken and the mass matrices of the fermions
and the gauge bosons are diagonalized, the magnetic moment-type couplings of
the leptons to gauge boson $Z$ and the photon $\gamma$ are
\begin{eqnarray}\label{lfvlag}
{\cal L}^{Z, \gamma}_{eff}&=&
eg^{Z, \gamma}
{{ \pmatrix{{\overline e}\cr {\overline \mu} \cr{\overline \tau} \cr}}}^T
 \left \{
(-\frac{1}{ 2 m_\tau})( i k_\nu \sigma^{\mu\nu}) S^{Z, \gamma}
\right .  \nonumber\\
&&\times U_l
 \left . \pmatrix{ 0 & & \cr
  & 0 & \cr
  & &  1 \cr }
  U_l^\dagger\right \}
\pmatrix{e \cr \mu \cr \tau \cr} V_{\mu}, \nonumber \\
 &&V_{\mu}=Z_{\mu}, A_{\mu},
\end{eqnarray}
where $g^Z =  1/(4s_Wc_W), g^\gamma=1$, and
\begin{eqnarray}
\label{szsg}
 S^Z&=&-g^Z \frac{2\sqrt 2}{e}\frac{m_\tau v}{\Lambda^2}\left [
   C_{\tau W }c_W-C_{\tau B}s_W\right ],\nonumber\\
\label{eq15}
 S^{\gamma}&=&\frac{2\sqrt 2}{e}\frac{m_\tau v}{\Lambda^2}\left [
   C_{\tau W }s_W-C_{\tau B }c_W \right ].
\end{eqnarray}

The matrix $U_l$ in Eq.~(\ref{lfvlag}) is the unitary matrix which
diagonalizes the mass matrix of the charged lepton.
In the SM, which corresponds to ${\cal L}_{eff}$ in the limit of $\Lambda
\to \infty$, the matrix $U_l$ is not measurable because of the zero neutrino
masses. Furthermore, the universality of the gauge interaction guarantees
the absence of the flavor changing neutral current and the lepton flavor violation
in the lepton sector.

The lagrangian ${\cal L}^{Z, \gamma}_{eff}$
in Eq. (\ref{lfvlag}) gives rise to the anomalous magnetic moment
of muon (g-2) and the rare decays $\tau \to \mu \gamma$
etc, whose relative size, however,
 depends on the rotation matrix $U_l$. In Ref.~\cite{tau}, we have
 shown that
$\mu \to e \gamma$ puts a very stringent constraint on the product of
the tau-electron and muon-electron mixing angles.
So in the following discussion we assume that
the tau mixes only with the muon, but not with the electron and for simplicity we have
assumed that rotation matrices of left-handed leptons and the right-handed
leptons are equal \footnote{We have also considered the case with different
left-handed and the right-handed rotation matrices and find our conclusions
unchanged.},
\bea  \label{matrix}
U_l=\pmatrix{ 1 & 0 & 0\cr
  0 & c & s \cr
  0 & -s & c \cr },
\eea
where $s=\sin\theta$ and $c=\cos\theta$ with $\theta$ being mixing angle.
The decay width of $\tau \to \mu \gamma$ is given by
\bea
\Gamma(\tau \to \mu \gamma)= \frac{\alpha}{8} m_{\tau} (s ~c~
S^\gamma)^2,
\eea
and the new contribution to the tau and muon anomalous magnetic moments are
given by
\bea
\vert \delta a_\tau \vert =
\left \vert c^2 S^{\gamma} \right \vert;~~
\vert \delta a_\mu \vert =
\left \vert s^2 S^{\gamma} \right \vert\frac{m_\mu}{m_\tau} .
\eea
%%%%%%%%%%%

Within the parameter space of $C_{\tau B}, ~C_{\tau W}, \Lambda $ and $\sin\theta$,
we have quantitatively studied the new physics effects on the
$\tau \to \mu \gamma $,
$Z \to \tau\mu $ , $\delta a_{\tau} $ and $\delta a_\mu$. We find that the predicted
$Br(\tau \to \mu \gamma)$ is correlated to $\delta a_\mu$, which is shown in Fig.
1. The figure also shows that  $Br(Z \to \tau \mu)$ could be quite
close to the experimental limit. On the other hand, given the
current limits on $Br(\tau \to \mu \gamma)$ and $\delta a_\mu$,
 the predicted $Br(Z \to \tau \mu)$ and $\delta a_\tau$ are well below the experimental limits.
  Numerically  to fit
$\delta a_{\mu}$ to be within 90\% C.L. of the new experimental
data, we obtain that
$s^2$ has to be bigger than 0.46, which is quite reasonable in some models~\cite{xing}.
 And $Br(\tau \to \mu \gamma)$ could be close to the current experimental limit.
For example, taking
$s^2\sim 0.5$, $C_{\tau B}, C_{\tau W} \sim 1/{(4 \pi)}^2 $
~\cite{wudka}
and $\Lambda \sim 17.5$ TeV, the new physics contribution to muon
(g-2) is,
\be
\delta a_\mu = 221 \times {10}^{-11},
\ee
which is within 90\% C.L. of the experimental data
\be
a^{exp}_\mu-a^{SM}_\mu = 426 \pm 165  \times {10}^{-11},
\ee
and $ Br( \tau \to \mu \gamma ) = 1 \times {10}^{-6} $
which is close to the current experimental limit
$ Br^{exp}( \tau \to \mu \gamma ) = 1.1 \times {10}^{-6} $~\cite{pdg}.
From Fig.1, one can see that increasing the mixing angle $\theta$
will decrease the value of predicted $Br(\tau \to \mu \gamma)$,
however for a large range of the mixing angle it remains to be
close to the current experimental limit.

\begin{figure}[thb]
\centerline{\epsfysize 2. truein \epsfbox{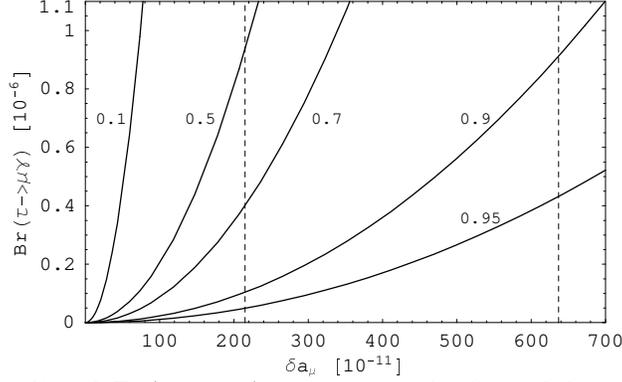}}
\caption[]{Plot of the predicted $Br(\tau \to \mu \gamma)$ versus $\delta a_\mu$:
 the five solid curves correspond to mixing angles
$\sin^2 \theta =$0.1, 0.5, 0.7, 0.9, 0.95, and between the
dashed lines is the
region of experimental value of $\delta a_\mu$ with 90\% C.L..
\label{fig1}}
\end{figure}

Now we consider a
phenomenological model where a heavy neutral vector boson $Z^\prime_\mu$
is introduced and coupled to only the tau lepton,
\be
{\cal L}^{int} = - g Z^\prime_\mu  {\bar \tau} \gamma^\mu \tau ,
\label{zprime}
\ee
where g is the coupling constant. In Eq.~(\ref{zprime}) we have considered
only the vectorial coupling for simplicity. The couplings of $Z^\prime$
to other fermions are not given since they are irrelevant for the purpose of
this paper.

After diagonalizing the lepton mass matrix by $U_l$ in Eq.~(\ref{matrix}),
Eq.~(\ref{zprime}) becomes
\bea
{\cal L}^{int}&=& - g Z^\prime_\mu[ c^2~\bar{\tau} \gamma^\mu \tau
+ s~c~\bar{\tau} \gamma^\mu \mu \nonumber\\
&&+ s~c~ \bar{\mu}\gamma^\mu \tau
+ s^2~\bar{\mu} \gamma^\mu \mu ].
\label {zpleps}
\eea
At one-loop level, the anomalous magnetic moment of the muon is generated as shown
in Fig.~2(a) and Fig.~2(b). However compared with Fig.~2(a), contribution
to muon (g-2) from Fig.~2(b) is less suppressed by the mixing angles
in the vertices and is enhanced by the tau lepton mass in the internal propagator,
which gives rise to
\bea
\delta a_\mu&=& \frac{g^2}{ { (4 \pi )}^2 }
\frac{ 8~s^2~c^2~ m_\tau m_\mu}{  M^2_{Z^\prime} }
I( \frac{ m^2_\tau }{  M^2_{Z^\prime} } ,
\frac{ m^2_\mu }{  M^2_{Z^\prime} } ),\nonumber\\
I &=& \int^1_0 \frac{y ( 1-y )  dy }{
 y^2  m^2_\mu / M^2_{Z^\prime}
- y (  1 - m^2_\tau / M^2_{Z^\prime} + m^2_\mu / M^2_{Z^\prime} )
+ 1 }    \nonumber \\
&=& \frac{1}{2} + {\cal O} ( \frac{ m^2_\tau }{  M^2_{Z^\prime} } ,
\frac{ m^2_\mu }{  M^2_{Z^\prime} } ).
\eea
%\bea
%\delta a_{\mu} =
%\frac{g^2}{(4 \pi )^2 }
%8 s^2
%\frac{ m^2_\tau m^2_\mu}{  m^2_{Z^\prime} }
%(\frac{1}{2} + O(
%\frac{ m^2_\tau }{ m^2_{Z^\prime} },
%\frac{ m^2_\mu }{  m^2_{Z^\prime} }
%) )
%\eea
Taking $ g^2 \le 4\pi $ to make the perturbative calculation
reliable and  $s^2 \sim 0.5$ to maximize $s^2c^2$,
we find that to fit muon (g-2) to be within 90\% C.L. of the
experimental data, $M_{ Z^\prime} $ has to be lighter than 2.6 TeV.
With these input parameters of $g, s$ and $M_{Z^\prime}$,
the predicted rate of $ \tau \to \mu \gamma $ is $9.3\times 10^{-7}$.
The result holds unchanged for weakly interaction theory if the
ratio of $g^2/M^2_{Z^\prime}$ remains the same as of $4\pi/(2.6
TeV)^2$. In this type of models, there are in general two sources
which contribute to $Z \to \tau \mu$. One is the radiative vertex
correction by the $Z^\prime$ vector boson with the result given by
\bea
Br(Z\to \tau \bar{\mu})=
Br(Z\to l \bar{l}) \times \left (\frac{g^2 s c^3}{12 \pi^2}~
\frac{M_Z^2}{M_{Z^{\prime}}^2}~ \ln \left(\frac{M_{Z^{\prime}}^2}{M_Z^2} \right)\right
)^2.
\eea
Taking $ g^2 \le 4\pi $, $s^2 \sim 0.5$ and $M_{ Z^\prime} =$ 2.6 TeV,
we find that the branching ratio of $Z\to \tau \bar{\mu}$ is about
$1.6\times 10^{-9}$, which is well below the experiment limit
$1.2\times 10^{-5}$ \cite{pdg}. If a $Z - Z^\prime$ mixing
$\theta_{Z^\prime}$ is explicitly introduced, we will have
\cite{lan}
\bea \label{2}
Br(Z\to \tau \bar{\mu})=
Br(Z\to l \bar{l}) \times \frac
{2 g^2 \sin^2{\theta_{Z^\prime}} s^2 c^2 }
{g_Z^2 (g_L^2+g_R^2)},
\eea
where $\theta_{Z^\prime}$ is the $Z$-$Z^{\prime}$ mixing angle,
$g_Z$, $g_{(L,R)}$ are the $Z$ coupling constants in the SM.
Given the experimental limit on $\theta_{Z^\prime}$, $Br(Z \to
\tau \mu)$ is quite safe with the experimental constraint
\cite{pdg}. For example,
taking  $ g^2 \le 4\pi $, $s^2 \sim 0.5$ and $\sin \theta_{Z^\prime} \sim
10^{-3}$,
Eq. (\ref{2}) shows that
the branching ratio of $Z\to \tau \bar{\mu}$ is about
$3\times 10^{-6}$.
A full study of the $ Z^\prime $ physics on the electroweak observables depends
on the detail of the model parameters.
This goes beyond the scope of this paper.

\begin{figure}[thb]
\centerline{\epsfysize 1.3 truein \epsfbox{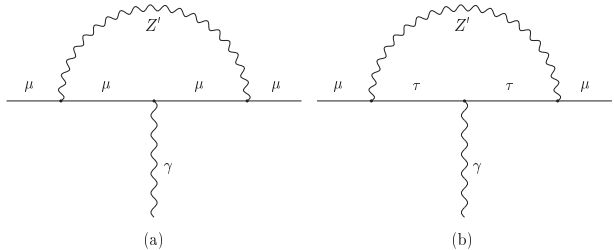}}
\caption[]{(a) One-loop $Z^\prime$ radiative correction to $a_\mu$;
(b) One-loop correction to $a_\mu$ from lepton flavor violating
couplings.
\label{fig2}}
\end{figure}

In summary, we have proposed in this paper a possible solution to
the discrepancy of the muon $( g-2 )$ between the standard model
prediction and the new experimental data,
by introducing the new physics with the third family. Compared to
other approaches to $\delta a_\mu$ \cite{new}, ours is the only
one which relates the new physics responsible for $\delta a_\mu$
to another exciting possibility, {\it i.e.} lepton flavor
violation. Furthermore,
in model with $Z^\prime$, our results show that $M_{Z^\prime} $ has to be lighter than 2.6 TeV.

{\bf Acknowledgments}:
This work was supported in part by National Natural Science Foundation of
China and by the Ministry of Science and Technology of China under
Grant No. NKBRSF G19990754.

\end{document}